\documentstyle[12pt,aps,prabib,psfig]{revtex}
\onecolumn 
\tighten
\begin{document}

\title{
\begin{flushright}
\vspace{-1cm}
{\normalsize MC/TH 97/09\\}
\vspace{1cm}
\end{flushright}
Disoriented chiral condensate formation from tubes\\
of hot quark plasma}

\author{Abdellatif Abada\thanks{Present address: BP International Limited, 
Britannic House, 1 Finsbury Circus, London, EC2M 7BA, England}
 and Michael C. Birse}

\address{ Theoretical Physics Group, Department of Physics and Astronomy,\\
University of Manchester, Manchester M13 9PL, England}

\maketitle

\bigskip
\bigskip
\begin{abstract}
We investigate the time evolution of a system of quarks interacting with 
$\sigma$ and pion fields starting from an initial configuration consisting of a
tube of hot quark plasma undergoing a boost-invariant longitudinal expansion.
We work within the framework of the linear sigma model using classical
transport equations for the quarks coupled to the mean-field equations for the
meson fields. In certain cases we find strong amplifications of any initial 
pion fields. For large-radius tubes, starting from quark densities that are
very close to critical, we find that a disoriented chiral condensate can form
in the centre of the tube. Eventually the collapse of the tube drives this
state back to the true vacuum. This process converts the disoriented
condensate, dominated by long-wavelength pion modes, into a coherent
excitation of the pion field that includes significant components with
transverse momenta of around 400 MeV. In contrast, for narrow tubes or larger
initial temperatures, amplification occurs only via the pion-laser-like
mechanism found previously for spherical systems. In addition, we find that
explicit chiral symmetry breaking significantly suppresses the formation of
disoriented condensates.
\end{abstract}

\section{Introduction}

Relativistic heavy-ion collisions provide a way to form regions of hot, dense
hadronic matter. If the temperature is high enough, this matter is expected to
be in a phase where chiral symmetry is restored and quarks are unconfined.
Recently there has been much interest in the possibility that, as such a
system cools, it could lead to regions in which the quark condensate is
misaligned with respect to the physical vacuum. These regions can also be
thought of as coherent excitations of the pion fields along particular
directions in isospin space. Such a state is known as a disoriented chiral
condensate (DCC) and a signal for its formation would be anomalously large
event-by-event fluctuations in the ratio of charged to neutral
pions.\footnote{For reviews of DCC's, as well as further references,
see:\cite{Raj,Bjo,BKrev,ARS,Raj2}.}

Many of the studies of DCC's found in the literature use idealised geometries
in order to simplify the calculations. These include uniform matter in a
finite box\cite{RW}, infinite matter undergoing a boost-invariant expansion in
one direction\cite{BK,HW,CM,CKMP,BCG,Ran1} and isotropically expanding infinite
matter\cite{GM,LDC,Ran1}. An important exception is the work of Asakawa {\it et
al.}\cite{AHW}, who considered the classical evolution of the chiral fields
for cylindrical systems undergoing boost-invariant longitudinal expansion.

In a previous paper\cite{AB} (hereafter referred to as I), we examined the
effects of finite size on the evolution of the chiral fields in the case
of spherical systems. The framework used was the linear sigma model\cite{GML}
with explicit quark degrees of freedom. This model has also been studied by
Csernai and Mishustin\cite{CM}, for the case of infinite matter expanding in
one direction.\footnote{Other recent work using this model can be found in
Refs.\cite{MS,BMFC,FMC}.} Here we extend our studies to the case of cylindrical
systems undergoing a boost-invariant longitudinal expansion. This geometry 
should be more relevant to ultra-relativistic heavy-ion collisions, and in
particular to the central regions of such collisions where particle production
is expected to be independent of rapidity\cite{Bj83}.

As in Ref.\cite{CM}, we assume a rapid quench that leaves the quarks out of 
thermal equilibrium with the chiral fields. The subsequent evolution of the
system is then described by the classical Euler-Lagrange equations for the
fields coupled to a relativistic transport equation for the quarks\cite{SR}.
We work in the classical limit where the transport equation reduces to
a relativistic Vlasov equation for the distribution of the quarks in phase 
space. Further details of the model and method of solution can be found in I.

For spherical droplets of hot quark matter, our studies in I showed that the
quarks stream rapidly away leaving the chiral fields in an unstable
configuration. During their subsequent evolution these fields always ``roll"
towards the physical vacuum. This behaviour can be thought of as the inward
collapse of the surface of the droplet as the quarks escape from it. Although
such systems show no tendency to form DCC's, we found that coherent
amplification of any initial pion fluctuation could occur through a
pion-laser-like mechanism. This is a consequence of the strong oscillations of
the $\sigma$ field that can pump energy into oscillations of the pion field.
Similar behaviour has also been seen in Refs.\cite{BVH,MM}. We have shown in I
that this mechanism is a robust one: for example, it does not require a chiral
phase transition in order to produce enhanced pion fields.

For cylindrical systems studied here, we find a competition between two
mechanisms. The first consists of the transverse flow of the quarks and 
resulting collapse of the surface of the tube. This leads to behaviour that is
similar to that seen in the spherical case. The second is the dilution of the 
quark density inside the tube as a result of its longitudinal expansion. The 
latter can lead to DCC formation, as one might expect from the results
of Ref.\cite{CM}. However, the extent to which this mechanism operates is very
sensitive to details of the initial configuration. The formation time for any
DCC relative to the time for the tube to collapse determines whether a DCC
forms, and how long it can exist. The collapse also changes the momentum
distribution of the pions from that of the initial DCC, by coherently exciting
pion modes with higher momenta. In addition we find that inclusion of explicit
chiral symmetry breaking, with the strength needed to give the observed pion
mass, significantly suppresses DCC formation.

The paper is organized as follows. In Sec.~II we present the linear sigma
model that we use together with the classical transport and field equations.
Since the basic approach has already been described in I, we concentrate on
those features that are specific to the longitudinally expanding cylinder. Our
results are described in Sec.~III and we discuss their implications in Sec.~IV.

\section{Model}

We work here with the linear sigma model\cite{GML}, which provides a simple
model for the physics associated with chiral symmetry. The model describes
quarks interacting with a chiral four-vector of meson fields $(\sigma,
\mbox{\boldmath $\pi$})$. Since we study here configurations in which only one
component of the pion field is nonvanishing, we keep only one pion field and,
as in I, we further simplify the model by neglecting the isospin dependence of
the quark-pion coupling. The Lagrangian we use is thus
\begin{equation}\label{Lagr}
{\cal L} = \bar\psi[i\partial\llap/-g(\sigma+i\pi\gamma_5)]\psi 
+\frac{1}{2}(\partial_{\mu}\sigma\partial^{\mu}\sigma 
+ \partial_{\mu}\pi\partial^{\mu}\pi) - U(\sigma,\pi),
\end{equation}
in which the meson fields of an O(2) linear sigma model are coupled to two
flavours and three colours of quark. 

The interactions among the meson fields are described by the potential $U$,
which we take to be of the form
\begin{equation}\label{mespot}
U(\sigma,\pi)= \frac{\lambda^2}{4}\bigl(\sigma^2+\pi^2-\nu^2
\bigr)^2-f_{\pi} m_{\pi}^2\sigma,
\end{equation}
where $f_\pi=93$ MeV is the pion decay constant. The ``Mexican-hat" form of
this potential leads to spontaneous breaking of the chiral symmetry through
the nonzero vacuum expectation value of $\sigma$, which corresponds to the
quark condensate of the QCD vacuum. The parameter $\lambda$ is chosen to give
the $\sigma$ meson a mass in the range 600--1000 MeV. In the physical vacuum
the quarks develop a mass $M_q=gf_\pi$. We consider values of $g$ that
correspond to quark masses in the range 300-500 MeV.

For the situation considered here, it is convenient to work in terms of polar
coordinates $\rho$ and $\phi$ for the transverse position, proper time
$\tau=\sqrt{t^2-z^2}$, and (space-time) rapidity $\eta={1\over
2}\ln[(t+z)/(t-z)]$. For a cylindrically symmetric system undergoing a
boost-invariant expansion, the fields depend only on $\rho$ and $\tau$. In this
case the equations for the meson fields take the form
\begin{equation}\label{sig}
{1\over\tau}{\partial\over \partial\tau}\left(\tau{\partial\sigma\over 
\partial\tau}\right)={1\over\rho}{\partial\over\partial\rho}\left(\rho
{\partial\sigma\over\partial\rho}\right)
-\left[\lambda^2\Bigl(\sigma^2(\tau,\rho)+\pi^2(\tau,\rho)-\nu^2\Bigr) 
+ g^2 {\cal S}_{\rm q}(\tau,\rho)\right]\sigma(\tau,\rho) + f_{\pi} m_{\pi}^2,
\end{equation}
\begin{equation}\label{pi}
{1\over\tau}{\partial\over \partial\tau}\left(\tau{\partial\pi\over 
\partial\tau}\right)={1\over\rho}{\partial\over\partial\rho}\left(\rho
{\partial\pi\over\partial\rho}\right)
-\left[\lambda^2\Bigl(\sigma^2(\tau,\rho)+\pi^2(\tau,\rho)-\nu^2\Bigr) 
+ g^2 {\cal S}_{\rm q}(\tau,\rho)\right]\pi(\tau,r).
\end{equation}
The source density in these equations, ${\cal S}_{\rm q}(\tau,\rho)$, is 
proportional to the scalar density of quarks. Its detailed form is given below.

The Vlasov equation\cite{SR} is conveniently expressed in terms of momentum
variables defined in the local rest frame of the matter. In this frame, which
is specified by the unit four-vector $u=(t,{\bf 0},z)/\tau$\cite{Bj83}, 
the longitudinal momentum $p_\parallel$ corresponding to a classical
particle with three-momentum {\bf p} moving in the presence of mean scalar and
pseudoscalar fields is
\begin{equation}
p_\parallel={p_z t-Ez\over\tau},
\end{equation}
where
\begin{equation}\label{quarkE}
E(x,{\bf p})=\sqrt{{\bf p}^2 + M^2(x)},
\end{equation}
and
\begin{equation}\label{quarkM}
M(x)=g\sqrt{\sigma^2(x)+\pi^2(x)}.
\end{equation}
Similarly, the energy of such a particle in this frame is
\begin{equation}
\epsilon=p\cdot u={Et-p_z z\over\tau}.
\end{equation}
The transverse components of the momentum, denoted ${\bf p}_\perp$, are of
course unchanged by the boost to the local rest frame. In terms of these
variables, the Vlasov equation for the phase-space distribution of quarks
$f(\tau,\eta,{\bf r}_\perp,p_\parallel,{\bf p}_\perp)$ can be written
\begin{equation}
\left[{\partial \over\partial\tau}
+{p_\parallel\over\tau\epsilon}{\partial\over\partial\eta}
+{1\over\epsilon}{\bf p}_\perp\cdot\nabla_{{\bf r}_\perp}
-{p_\parallel\over\tau}{\partial \over\partial p_\parallel}
-\Bigl(\nabla_{{\bf r}_\perp}\epsilon(\tau,{\bf r}_\perp,p_\parallel,
{\bf p}_\perp)\Bigr)\cdot\nabla_{{\bf p}_\perp}\right]
f(\tau,\eta,{\bf r}_\perp,p_\parallel,{\bf p}_\perp)=0,\label{Vl}
\end{equation}
where we have assumed boost invariance, {\it i.e.}~that $M(x)$ is independent
of $\eta$. The antiquark distribution, denoted $\tilde f (\tau,\eta,{\bf
r}_\perp,p_\parallel,{\bf p}_\perp)$, satisfies an equation of similar form.

The Vlasov equation (\ref{Vl}) describes freely streaming classical quarks and
antiquarks. These particles have energies that are related to their
three-momenta by Eq.~(\ref{quarkE}) and obey relativistic single-particle
equations of motion, which, in terms of the coordinates and momenta defined
above, take the form:
\begin{equation}
\dot{\bf r}_\perp(\tau)={{\bf p}_\perp(\tau)\over \epsilon},\label{spertr}
\end{equation}
\begin{equation}
\dot{\bf p}_\perp(\tau)=-\nabla_{{\bf r}_\perp}\epsilon,\label{speptr}
\end{equation}
\begin{equation}
\dot\eta(\tau)={p_\parallel(\tau)\over\tau\epsilon},\label{speeta}
\end{equation}
\begin{equation}
\dot p_\parallel(\tau)=-{p_\parallel(\tau)\over\tau},\label{speplong}
\end{equation}
where the overdots denote derivatives with respect to $\tau$ and the
particle's energy in the local rest frame is
\begin{equation}
\epsilon\Bigl(\tau,{\bf r}_\perp(\tau),p_\parallel(\tau),
{\bf p}_\perp(\tau)\Bigr)=\sqrt{M^2\Bigl(\tau,{\bf r}_\perp(\tau)\Bigr)
+p^2_\parallel(\tau)+{\bf p}^2_\perp(\tau)}.
\end{equation}

The source term in the field equations, $S_q(x)$, is given by the integral 
of $[f(x,{\bf p})+\tilde f(x,{\bf p})]/E(x,{\bf p})$ over all 
three-momenta\cite{SR,AB}. Changing variables to $p_\parallel$ and {\bf 
p}$_\perp$, we can write it in the form
\begin{equation}
S_q(x) = \int dp_\parallel\,d^2{\bf p}_\perp\, 
\frac{f(x,p_\parallel,{\bf p}_\perp)+\tilde f(x,p_\parallel,{\bf p}_\perp)}
{\epsilon(x,p_\parallel,{\bf p}_\perp)}.
\end{equation}

Rather than solving Eq.~(\ref{Vl}) directly as a partial differential equation
in seven dimensions, we use the test-particle method\cite{Won,SG}. In this
approach, the smooth distributions $f$ and $\tilde f$ are approximated by a
set of classical particles obeying the equations of motion
(\ref{spertr})--(\ref{speplong}). For the numerical results presented here, we
used 40~000 test quarks and antiquarks.

The numerical techniques for solving the equations of motion (\ref{sig}),
(\ref{pi}) and (\ref{Vl}) are very similar to those applied to a soliton bag
model in Ref.\cite{VBM}. Details of the method for the case of the linear sigma
model can be found in I. In the present case, the boost invariance of the
system leads to a number of simplifications. The quark distributions are
independent of $\eta$ and hence we do not need to consider the longitudinal
motion of the particles, Eq.~(\ref{speeta}). In addition, the longitudinal
momenta satisfy Eq.~(\ref{speplong}) and so they simply scale like
$1/\tau$\cite{CM}.

Because of the longitudinal expansion, there is no finite, conserved energy for
systems with this geometry. Nonetheless it is convenient to define an energy
per unit rapidity in the local rest frame. This takes the form
\begin{eqnarray}
E=\tau\int\!d^2{\bf r}_\perp\,\Biggl\{&&{1\over 2}\dot\sigma^2
+{1\over 2}\dot\pi^2+{1\over 2}(\nabla_{{\bf r}_\perp}\sigma)^2
+{1\over 2}(\nabla_{{\bf r}_\perp}\pi)^2+U(\sigma,\pi)\nonumber\\
&&+\int\!dp_\parallel\,d^2{\bf p}_\perp\,
\epsilon\left[f(\tau,\eta,{\bf r}_\perp,p_\parallel,{\bf p}_\perp)
+\tilde f(\tau,\eta,{\bf r}_\perp,p_\parallel,{\bf p}_\perp)\right]\Biggr\},
\end{eqnarray}
where a constant has been added to $U(\sigma,\pi)$ so that it vanishes in the
vacuum. By making use of the equations of motion, one finds that the rate of
change of $E$ is given by
\begin{eqnarray}
{dE\over d\tau}=-\int\!d^2{\bf r}_\perp\,\Biggl\{&&{1\over 2}
\dot\sigma^2+{1\over 2}\dot\pi^2-{1\over 2}(\nabla_{{\bf r}_\perp}\sigma)^2
-{1\over 2}(\nabla_{{\bf r}_\perp}\pi)^2-U(\sigma,\pi)\nonumber\\
&&+\int\!dp_\parallel\,d^2{\bf p}_\perp\,{p_\parallel^2\over \epsilon}
\left[f(\tau,\eta,{\bf r}_\perp,p_\parallel,{\bf p}_\perp)
+\tilde f(\tau,\eta,{\bf r}_\perp,p_\parallel,{\bf p}_\perp)\right]\Biggr\},
\end{eqnarray}
We have used this equation as a check on our numerical integration of the 
equations of motion.

\section{Results}

In this section, we present the results of our simulations corresponding to
various initial conditions of the system. As in I, the results shown are for
our ``standard" choice of parameters, $m_{\sigma}=1000$ MeV, $M_{\rm q}=300$
MeV and, if chiral-symmetry breaking is included, $m_{\pi}= 139$ MeV. For these
parameters, the temperature below which the phase with $\sigma\neq 0$ becomes 
the ground state is $T_0\simeq235$ MeV.

The initial conditions are specified by three parameters: the radius $r_0$ of
the tube, the temperature $T$ of the quark plasma and the proper time $\tau_0$
at which the evolution according to Eqs.~(\ref{sig}), (\ref{pi}) and
(\ref{Vl}) starts. A chemical potential $\mu$ can also be introduced to allow
for a nonzero initial baryon number, but this does not qualitatively change 
the behaviour of the system and so we present here only cases with $\mu=0$.
We assume, as before, that the plasma inside the tube is initially 
uniform.\footnote{For further details of the implementation of the initial
conditions in this model, see Sec.~IV of I.}

In order to study whether initial pionic fields can be amplified during the
evolution, we add a small pionic perturbation to the initial configuration. As
in our previous work, we have considered only uniform initial fluctuations of
the pion field since our aim is to study whether DCC formation is possible in
these systems. A more realistic approach would be to take initial fluctuations
from a thermal distribution using a method similar to that in\cite{Ran2}. We
would then expect to find the formation of regions of differently oriented
DCC's, as seen in the work of Asakawa {\it et al.}\cite{AHW}. Such studies
will require integration of the full three-dimensional equations of motion and
so will be much more computationally intensive.

For many choices of initial conditions, and in particular for cases with
relatively small initial tubes ($r_0<5$ fm), we find similar behaviour to that
seen for the spherical droplets of quark plasma studied in I. In these cases,
the rapid outward streaming of the quarks leaves the chiral fields in an
unstable configuration. The surface of the tube collapses inwards, with the
chiral fields ``rolling" towards the physical vacuum. The $\sigma$ field then
executes strong oscillations abouts its vacuum value. If a nonzero initial
pion field is present, then this can be amplified by the laser-like mechanism
also seen in I.

\begin{figure}[t]
\centerline{\psfig{figure=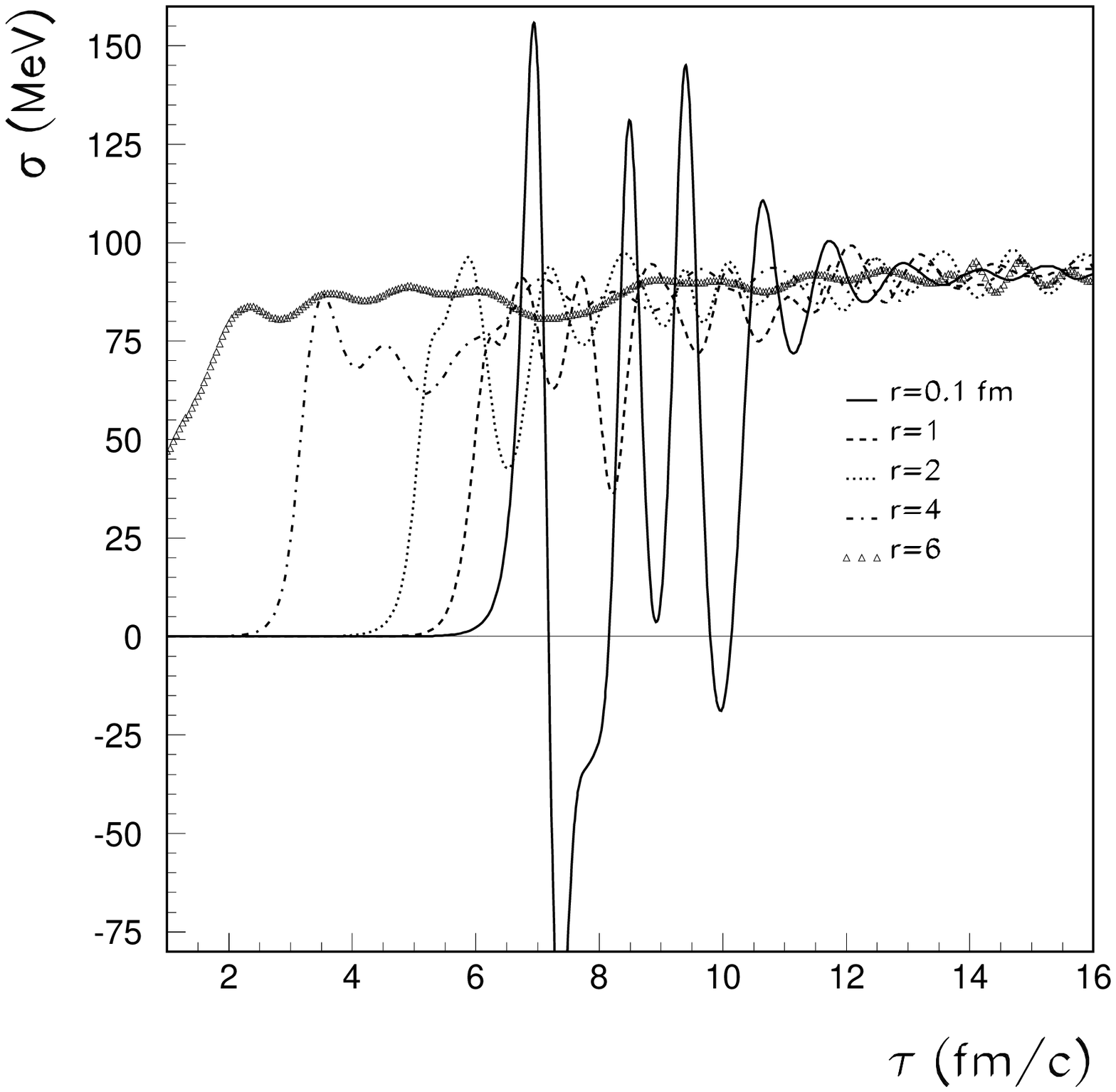}}
\caption{Time evolution of the $\sigma$ field at various radii. The parameters
of the model are $m_\sigma=1000$ MeV and $M_q=300$ MeV, and the system is
chirally symmetric ($m_\pi=0$). The tube has initial radius $r_0=6$ fm and
temperature $T=250$ MeV at $\tau_0=1$ fm/$c$.}
\label{figsig}
\end{figure}

\begin{figure}[t]
\centerline{\psfig{figure=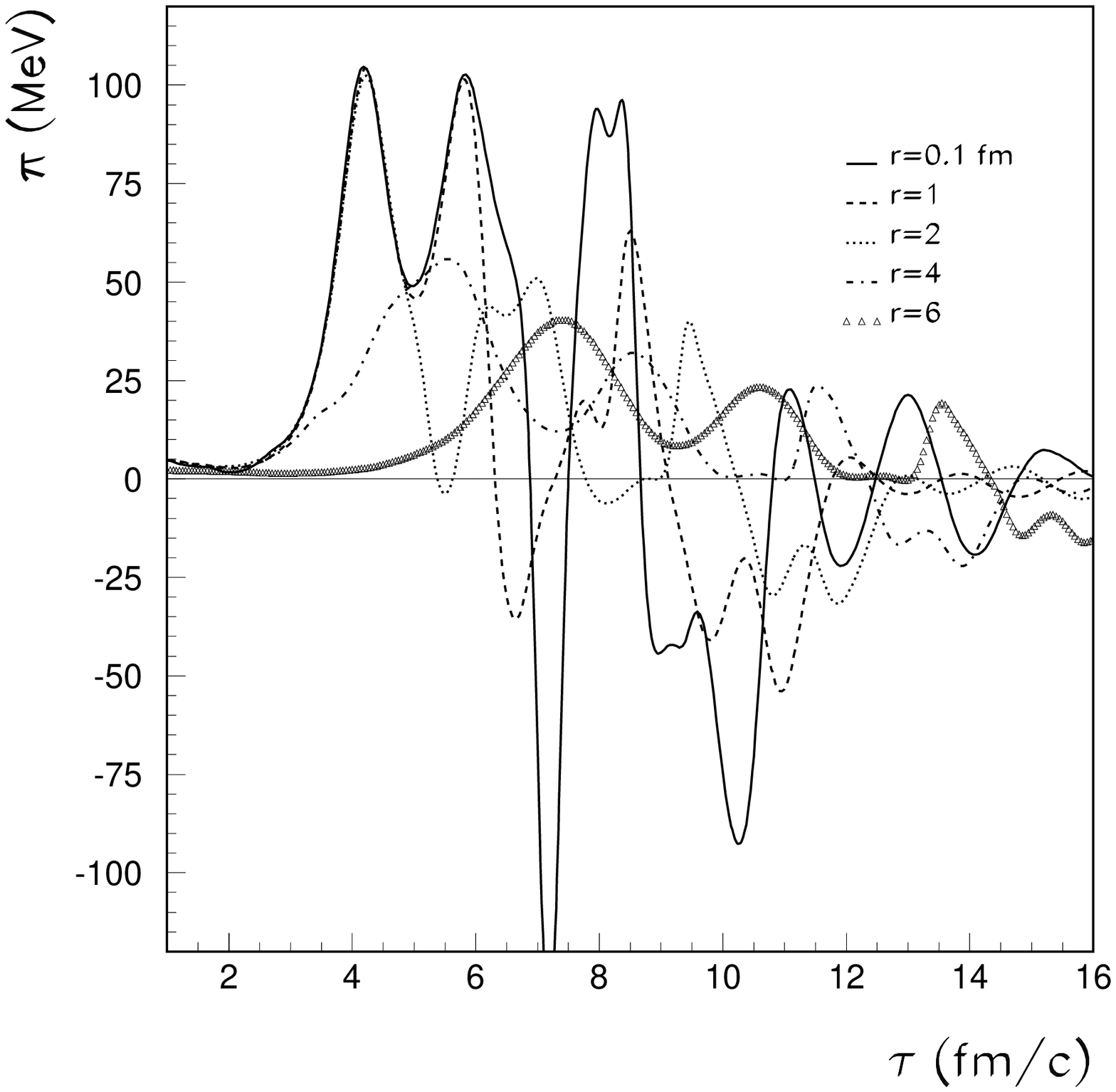}}
\caption{Time evolution of the pion field at various radii. The model
parameters and initial conditions are the same as in Fig.~1.}
\label{figpi}
\end{figure}

\begin{figure}[t]
\centerline{\psfig{figure=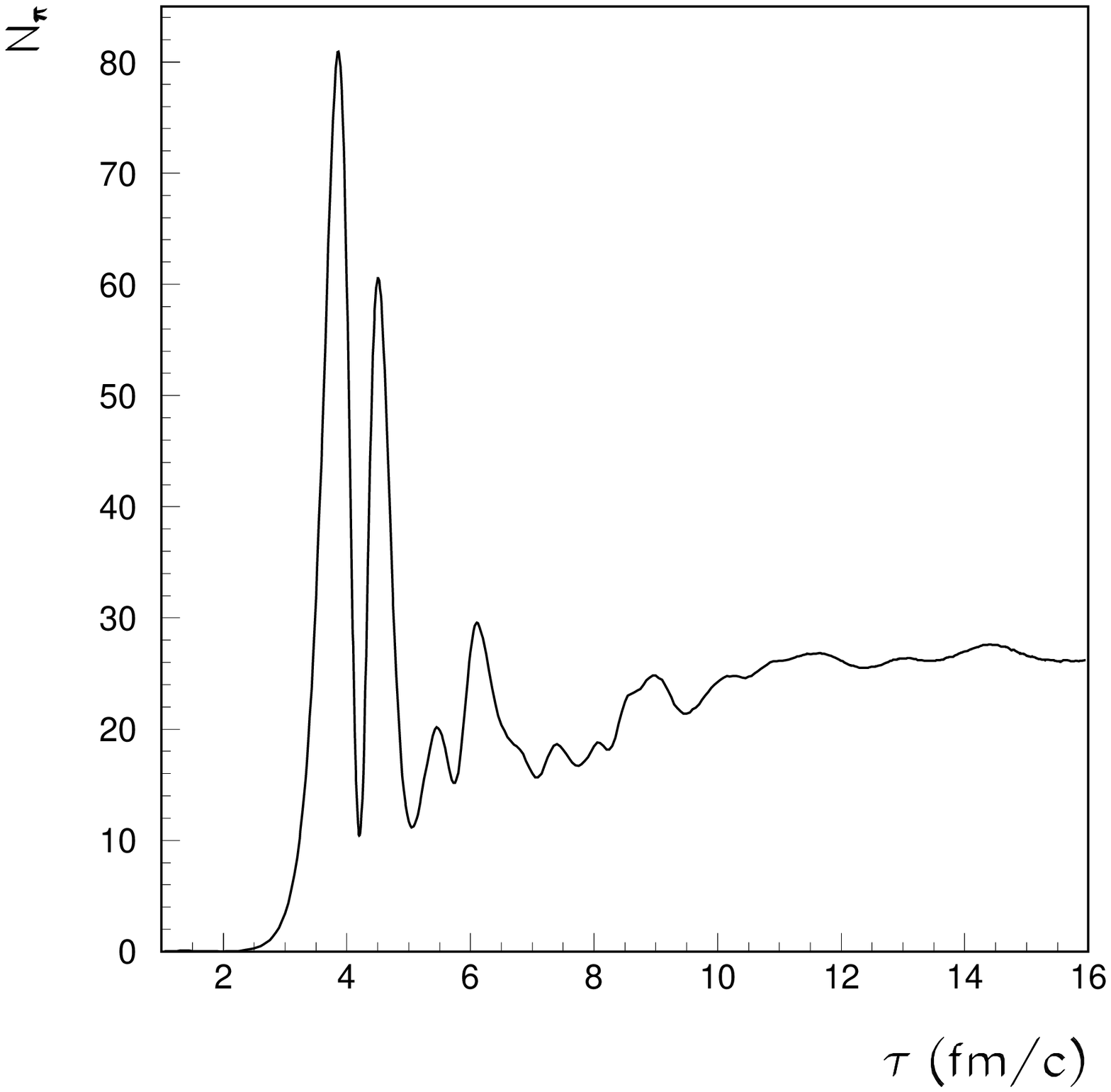}}
\caption{Time evolution of $N_\pi$. The model parameters and initial
conditions are the same as in Fig.~1.}
\label{fignpi}
\end{figure}

\begin{figure}[t]
\centerline{\psfig{figure=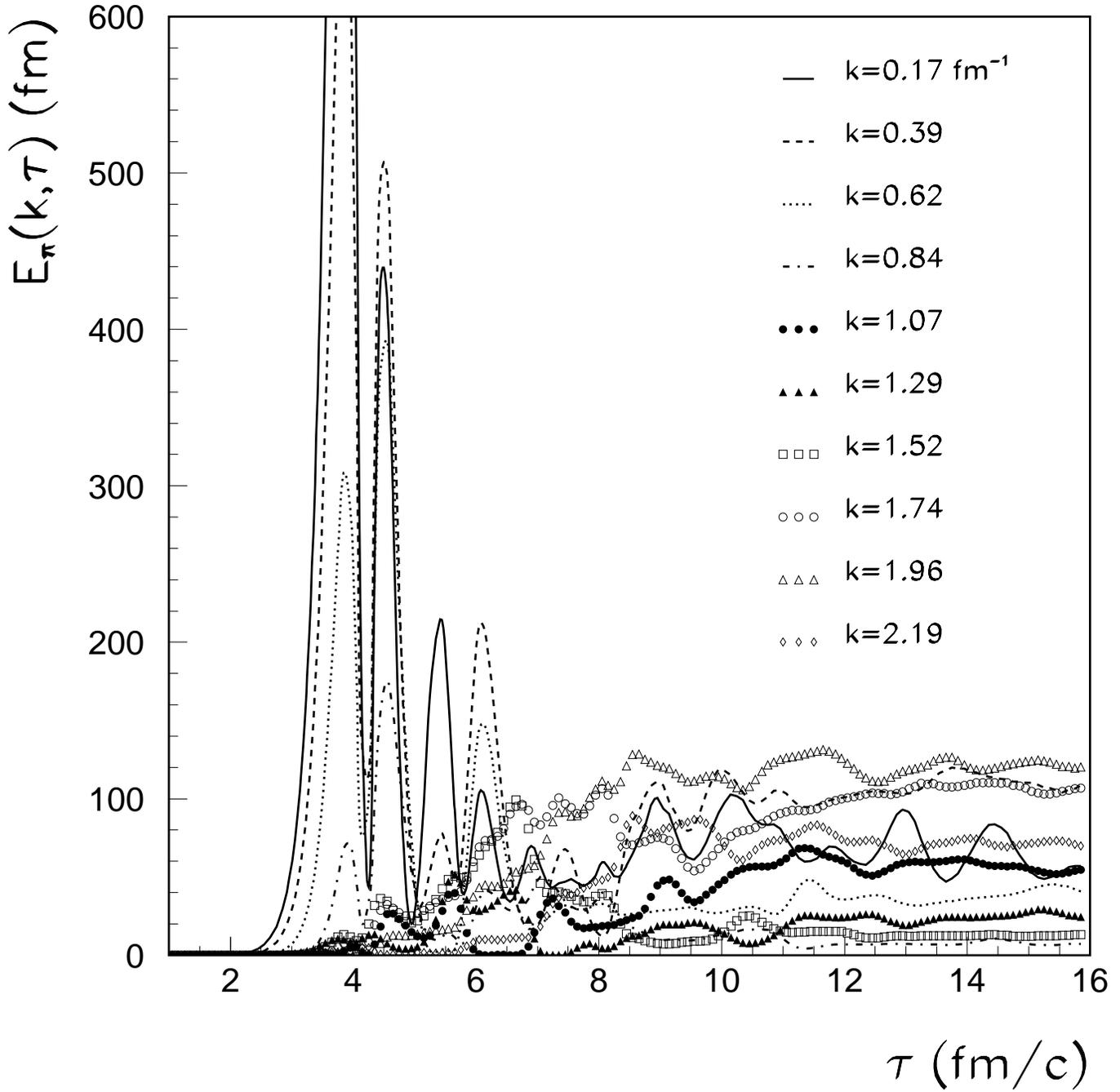}}
\caption{Time evolution of the pion energy density in momentum space for the
lowest ten modes of the field. The model parameters and initial conditions are
the same as in Fig.~1.}
\label{figtspec}
\end{figure}

\begin{figure}[t]
\centerline{\psfig{figure=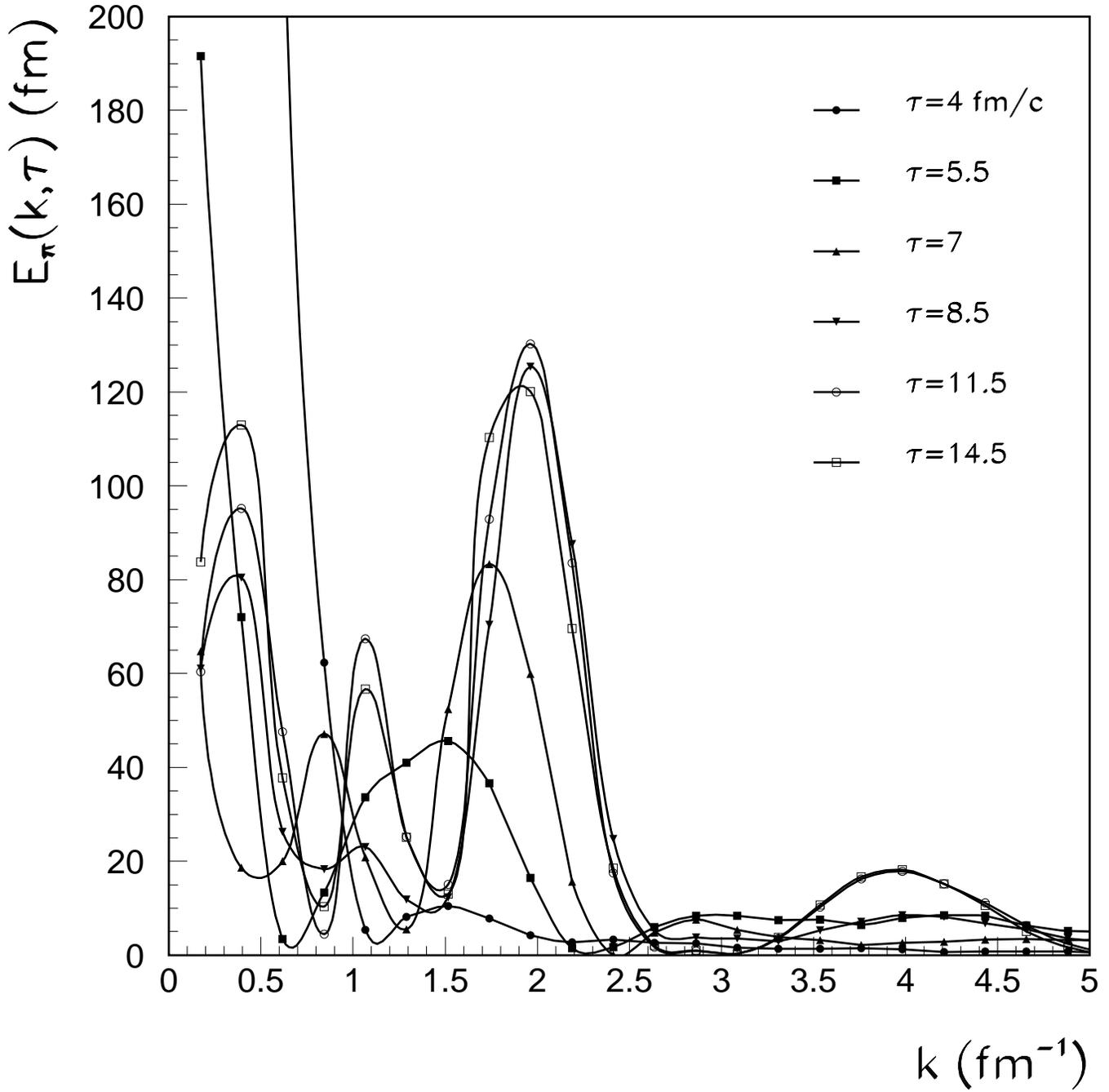}}
\caption{Momentum dependence of the pion energy density at various proper
times. The model parameters and initial conditions are the same as in Fig.~1.}
\label{figkspec}
\end{figure}

In contrast, for large enough initial tubes, we find a rather different
behaviour. This is particularly clear in the chiral limit. An example of this
type is shown in Figs.~\ref{figsig}--\ref{figkspec}. These show the behaviour
of a tube of initial radius $r_0=6$ fm and temperature $T=250$ MeV at
$\tau_0=1$ fm/$c$. This initial proper time is the same as that used in
Refs.\cite{HW,CKMP,BCG,LDC,AHW}, but is significantly smaller than the
estimates of the freeze-out time in the work of Csernai, Mishustin and
coworkers\cite{CM,MS,FMC}. Note that the two examples for which we display
results here are not necessarily the most realistic ones; rather they have
been chosen since they most clearly illustrate the types of behaviour that are
possible, and the conditions under which these can occur.

Figs.~\ref{figsig} and \ref{figpi} show the evolution of the chiral fields for
this example of large tube with an initial temperature just above $T_0$. The
behaviour of the central region of the tube is dominated, at least initially,
by the longitudinal expansion. In particular the quark density there drops
below its critical value before the quarks have a chance to start streaming
outwards. At this point the chirally restored phase becomes unstable and any
small fluctuation from it can start to grow exponentially. In the present case,
this happens at $\tau\sim 2.5$ fm/$c$, when the pion field at small radii can
be seen to rise rapidly and then oscillate about a value close to $f_\pi$.

This pion field is uniform across the central region of the tube. It is an
example of a DCC: a region of misaligned vacuum. The collapse of the surface
of the tube means that the DCC does not persist in this form for proper times
longer than $r_0/c$. In Figs.~\ref{figsig} and \ref{figpi}, one can see nonzero
$\sigma$ fields appearing at successively smaller radii and the corresponding
pion fields ceasing to oscillate in phase with that at the centre of the tube.
From $\tau\sim 7$ fm/$c$ onwards, the behaviour of the system resembles that
of the ones studied in I, the sigma field oscillating violently until
$\tau\sim 12$ fm/$c$ and then settling down to its vacuum value.

By Fourier analysing the pion field at successive times, we have also
investigated the spectrum of the pion modes that are excited. In terms of the 
transverse Fourier transform of the field,
\begin{equation}
\tilde \pi(\tau,{\bf k}_\perp)=\int\!d^2{\bf r}_\perp\,
\exp(i{\bf k}_\perp\cdot{\bf r}_\perp)\pi(\tau,\rho),
\end{equation}
and its proper-time derivative
$\dot{\tilde\pi}(\tau,{\bf k}_\perp)$
we define the corresponding intensity in momentum space (per unit rapidity):
\begin{equation} \label{Epi}
{\cal E}_{\pi}(t,{\bf k}_\perp)={\textstyle{\tau\over 2}}\Bigl(
\vert\dot{\tilde\pi}(t,{\bf k}_\perp)\vert^2 + \omega_k^2 \vert\tilde 
\pi(t,{\bf k}_\perp)\vert^2\Bigr),
\end{equation}
where $\omega_k=\sqrt{k_\perp^2+m_\pi^2}$. At large times, when the pion
fields are sufficiently weak that they are well described by a linearised
equation of motion, this is just the energy density of the pion field in
momentum space. A convenient measure of the total strength of the pion field
is provided by
\begin{equation}\label{Npi}
N_{\pi}(\tau)= \int\!\frac{{\rm d}^2 {\bf k}_\perp}{(2\pi)^2}\,{1\over
\omega_k}{\cal E}_{\pi}(t,{\bf k}_\perp),
\end{equation}
which, for weak fields, is equal to the total number of pions per unit
rapidity.

In Fig.~\ref{fignpi} we show the behaviour of $N_{\pi}(\tau)$. Initially this
rises very rapidly and then shows strong oscillations. This is similar to what
was seen in the studies of Rajagopal and Wilczek\cite{RW,Raj}. After $\tau\sim
7$ fm/$c$, these oscillations disappear and are replaced by a more gradual rise
in $N_{\pi}(\tau)$. From the spectrum shown in Fig.~\ref{figtspec}, we see that
the initial oscillations are due to the strong excitation of modes with the
lowest frequencies and so are characteristic of a DCC. As the tube collapses,
the nearly uniform DCC is destroyed and strength is removed from the lowest
modes. Nonetheless significant coherent pion fields continue to be present
although, unlike the original DCC, these oscillate in space and time. 

In Fig.~\ref{figkspec} we plot the same spectrum as a function of pion
momentum at various proper times. At $\tau=4$ fm/$c$ one sees the very large 
strength concentrated in modes with transverse momenta of less than 150 MeV.
This pattern is characteristic of the initial, nearly uniform DCC. At later
times, this strength falls off and modes with higher momenta become excited. In
particular, significant strength builds up in modes with transverse momenta in
the range 300--500 MeV. There is also a smaller peak for momenta of 700--900
MeV Although the final pion fields are no longer dominated by the lowest
momentum components, their amplitude still reflects the fact that a DCC was
formed. For example, $N_{\pi}$ is enhanced by a factor of about 2000. This is
far larger than the effects seen in the spherical systems studied in I, where
the maximum enhancement factors were less than 100.

\begin{figure}[t]
\centerline{\psfig{figure=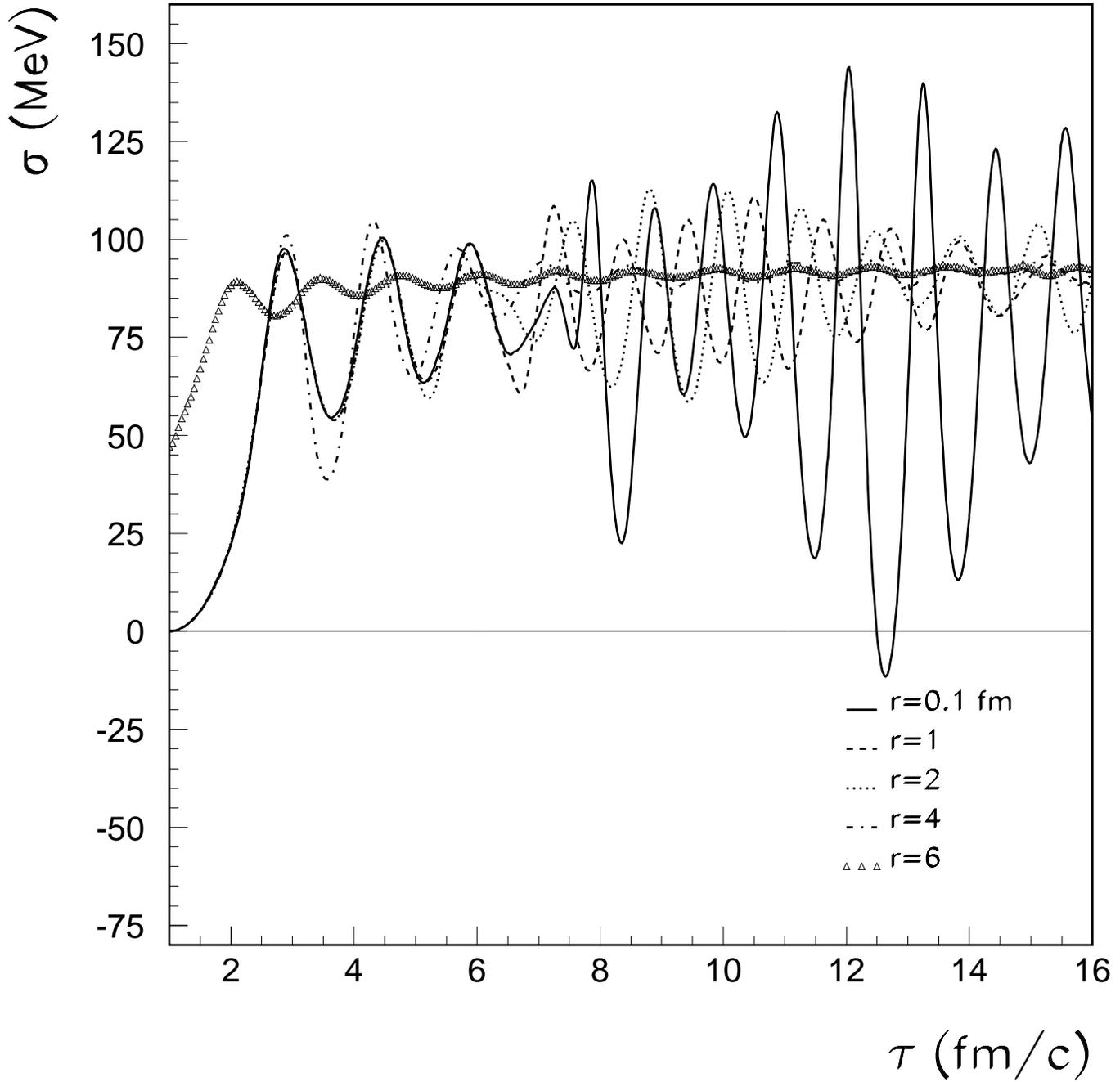}}
\caption{Time evolution of the $\sigma$ field at various radii. The parameters
of the model are $m_\sigma=1000$ MeV, $m_\pi=139$ MeV and $M_q=300$ MeV. The
tube has initial radius $r_0=6$ fm and temperature $T=225$ MeV at $\tau_0=1$
fm/$c$.}
\label{figsigm}
\end{figure}

\begin{figure}[t]
\centerline{\psfig{figure=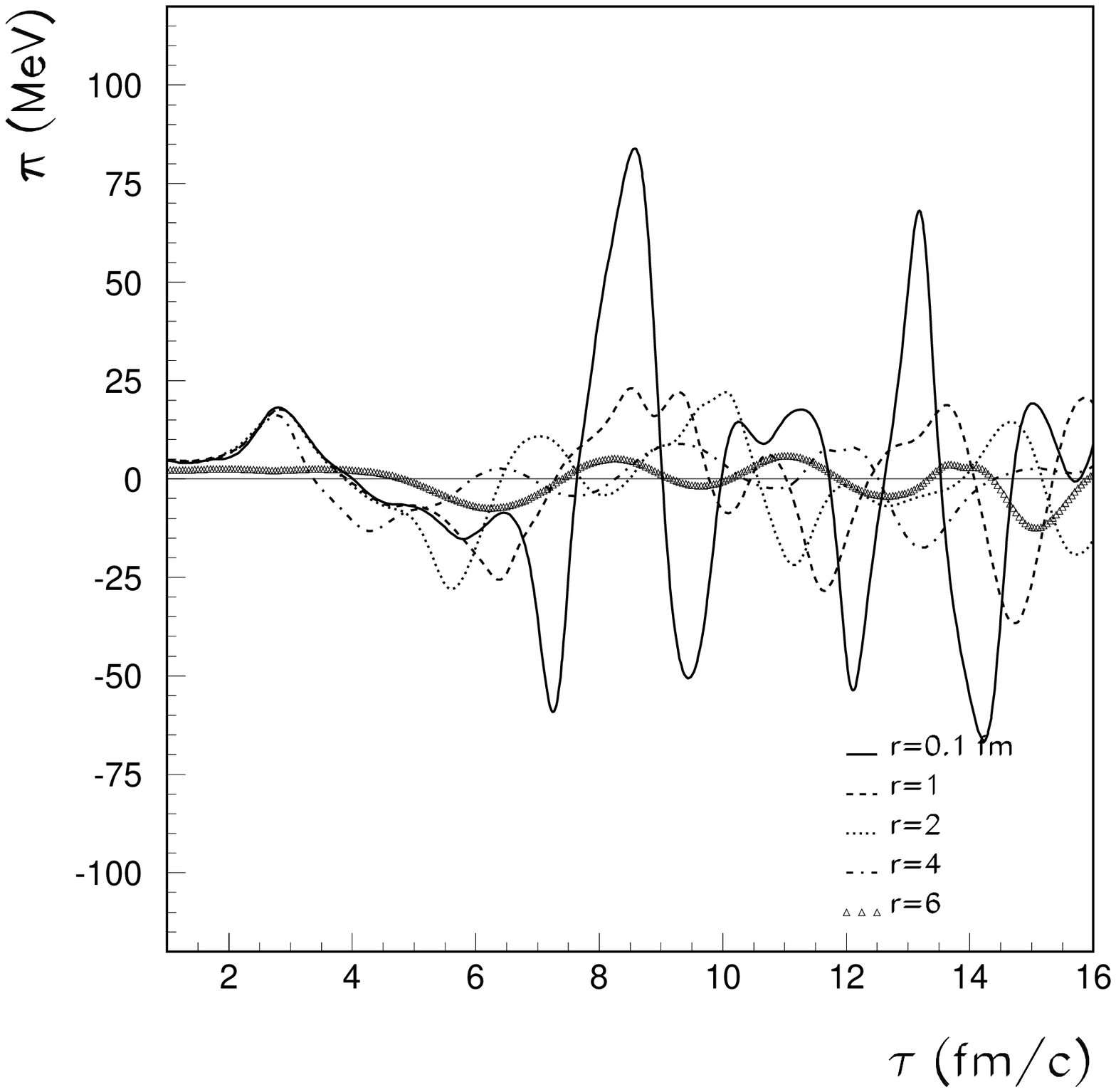}}
\caption{Time evolution of the pion field at various radii. The model
parameters and initial conditions are the same as in Fig.~4.}
\label{figpim}
\end{figure}

\begin{figure}[t]
\centerline{\psfig{figure=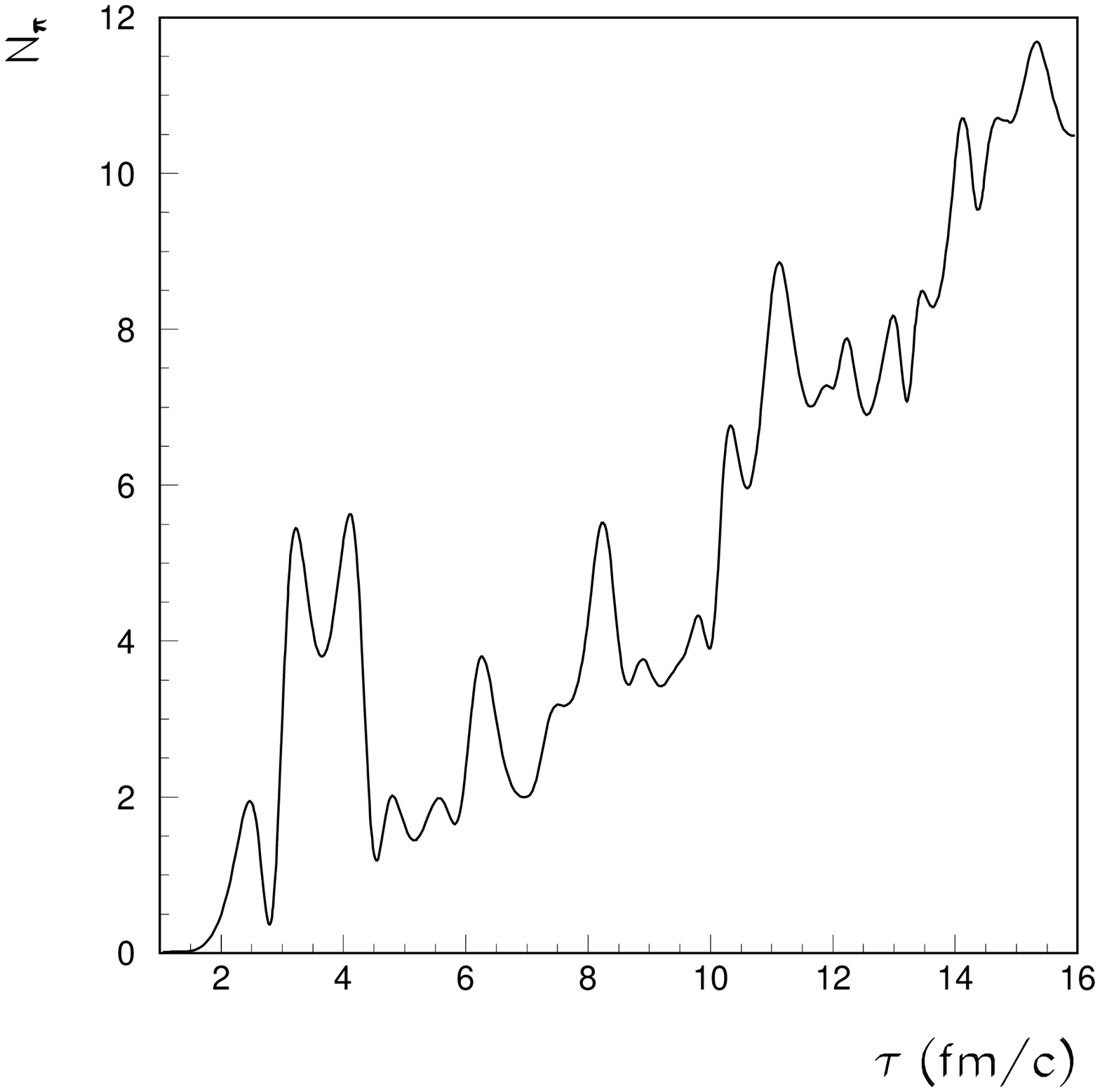}}
\caption{Time evolution of $N_\pi$. The model parameters and initial
conditions are the same as in Fig.~4.}
\label{fignpim}
\end{figure}

When explicit chiral symmetry breaking is present, in the form of the final 
term in the potential (\ref{mespot}), there is no phase transition but only
a crossover between vacua with small and large values of the $\sigma$ field.
For the physical value of $m_\pi$, the crossover is fairly rapid and so some of
the features of the phase transition do survive. Nonetheless explicit symmetry
breaking does lead to some qualitative differences in the evolution of the
chiral fields after a quench. An example is shown in
Figs.~\ref{figsigm}--\ref{fignpim}. In this case the radius of the initial
tube is again $r_0=6$ fm and $\tau_0=1$ fm. The initial temperature was taken
to be $T=225$ MeV, which lies within the crossover region for our standard
parameter set.

The explicit symmetry-breaking term in (\ref{mespot}) tilts the Mexican hat
potential in the direction of the true vacuum. Hence the system never evolves 
towards a maximally misaligned vacuum of the sort seen in the previous example.
Despite this, significant amplification of any initial pion field can occur and
so a DCC can still form. This is shown in Fig.~\ref{figpim} by the appearance
of a more-or-less uniform pion field of $\sim 20$ MeV in the centre of the
tube. This field starts to oscillate around the true vacuum, $\pi=0$. In an
infinite system, like those studied in Refs.\cite{HW,FMC}, these
oscillations would continue indefinitely. In the present case, they are
terminated at $\tau\simeq 7$ fm/$c$ by the collapse of the tube.

The fact that the fields inside the tube tend to roll towards the true vacuum,
as a result of the tilted potential, also means that much less energy is
released when the tube finally collapses. Indeed, in the example shown, the 
tube never completely collapses and a cylindrical region of cold quark matter
is formed. Formation of such a region of quark matter was predicted by Csernai
and Mishustin\cite{CM} and was also seen in I for parameter sets with strong
quark-meson couplings. This matter is created in a highly excited
configuration, as can be seen in Fig.~6 from the strong oscillations of the
$\sigma$ field from $\tau\sim 10$ fm/$c$ onwards. As this matter settles down,
it radiates a significant fraction of its energy in the form of pions, which
is the reason for the continued growth of $N_\pi$ with proper time seen in
Fig.~\ref{fignpim}. However, as discussed in I, we believe that the formation
of such matter represents an artifact of the model and so this behaviour should
not be taken too seriously. By $\tau\sim 10$ fm/$c$, before this behaviour
sets in, $N_\pi$ has been enhanced by a factor of about 200. Although this
enhancement is significantly less than that in the previous example, it is
still much larger than that found in the spherical systems studied in I.

\section{Discussion}

We have studied the evolution of systems of quarks coupled to chiral fields,
starting from an initial tube of hot quark plasma undergoing a boost-invariant
longitudinal expansion. Such a geometry is expected to be relevant to the 
central regions of ultra-relativistic heavy-ion collisions. We assume that a 
rapid quench occurs at some proper time $\tau_0$, leaving the quarks out of
thermal equilibrium with the chiral fields. The subsequent evolution of the
system is treated in the classical approximation, the quarks being described by
a relativistic Vlasov equation in the presence of $\sigma$ and pion fields
which satisfy the mean-field equations of the linear sigma model.

Unlike the spherical droplets studied previously in I, these cylindrical
configurations can form DCC's of the sort seen in infinite systems following
a quench\cite{RW}. Similar behaviour is also seen in infinite systems 
undergoing longitudinal expansion\cite{BK,CM,CKMP,BCG,AHW,FMC}. However we
find that the transverse flow of the quarks out of the tube tends to act
against the formation of a DCC. This effect does not occur for infinite
systems. It leads to an inward collapse of the surface of the tube, which is
similar to what happens for spherical droplets. Although this collapse does
not generate DCC's, it can enhance any initial pion fluctuation through a
pion-laser-like mechanism whereby strong oscillations of the $\sigma$ field
pump energy into the pion fields\cite{BVH,MM}.

In the cylindrical case, the presence of two competing mechanisms leads to
considerable sensitivity to the initial conditions. We find that significant 
DCC formation occurs only in large tubes ($r_0>5$ fm) where the longitudinal
expansion rapidly reduces the quark density below the critical value needed to
keep the system in the chirally symmetric phase. For a DCC to be produced, the
initial density of quarks and antiquarks should be close to the critical value
({\it i.e.}~the temperature should be close to $T_0$) and the proper time at
freeze-out should be small compared to the radius of the tube. 

If the initial proper time $\tau_0$ is taken to be $\sim 1$ fm/$c$, as in
Refs.\cite{HW,CKMP,BCG,LDC,AHW} and the illustrative examples shown here, then
tubes with radii of $\sim 5$ fm or more have time to form significant DCC's.
On the other hand, if $\tau_0$ is $\sim 5-10$ fm/$c$, as suggested by Csernai
and Mishustin\cite{CM,MS,FMC}, then it seems unlikely that DCC's can form in
tubes of realistic radii. Clearly it is crucial to have better models for the
early stages of ultra-relativistic heavy-ion collisions in order to determine
the temperature and proper time at freeze-out.

We also find that the subsequent evolution of such systems can significantly
affect the transverse-momentum distribution of the emitted pions. Even if a
system does form an initial DCC involving only very low-momentum modes of the
pion field, the energy released by the oscillations of the $\sigma$ field as
the tube collapses can lead to coherent excitation of pionic modes with higher
transverse momenta.

A further feature of our results is that explicit breaking of chiral symmetry
has a significant effect on the evolution of these systems. In particular, by
tilting the Mexican hat in the direction of the true vacuum, it reduces the
degree of misalignment of the vacuum that can appear at the centre of the
tube. A similar effect is also seen in infinite systems\cite{FMC}. As a result,
the enhancement of the number of pions, as given by Eq.~(\ref{Npi}), is
typically much smaller than that seen in the chiral limit. Nonetheless, even
in the broken-symmetry example shown above, the pionic enhancement is much
larger than that seen in the spherical systems studied in I, where no DCC was
formed.

\acknowledgements{We are grateful to G. Amelino-Camelia, J. D. Bjorken and J.
McGovern for helpful discussions, and to J. McGovern for a critical reading of
the manuscript. This work was supported by the EPSRC.}

\newpage

\newpage

\newpage

\newpage

\newpage

\newpage

\newpage

\newpage

\end{document}